\documentclass[twocolumn,showpacs,amsmath,amssymb]{revtex4}
\usepackage[dvips]{graphicx}% Include figure files
\usepackage{dcolumn}% Align table columns on decimal point
\usepackage{bm}% bold math

\begin{document}

\title{Laser-driven Sisyphus cooling in an optical dipole trap}

\author{Vladyslav V. Ivanov}
\email{vladivanov78@gmail.com}
\affiliation{Physics Department, University of Wisconsin,
Madison WI 53706}

\author{Subhadeep Gupta}
\affiliation{Physics Department, University of Washington,
Seattle WA 98195}
\homepage{www.phys.washington.edu/users/deepg/}

\date{\today}% It is always \today, today,

\begin{abstract}

We propose a laser-driven Sisyphus cooling scheme for atoms confined in a far off resonance optical dipole trap. Utilizing the differential trap-induced ac Stark shift, two electronic levels of the atom are resonantly coupled by a cooling laser preferentially near the trap bottom. After absorption of a cooling photon, the atom loses energy by climbing the steeper potential, and then spontaneously decays preferentially away from the trap bottom. The proposed method is particularly suited to cooling alkaline-earth-like atoms where two-level systems with narrow electronic transitions are present. Numerical simulations for the cases of $^{88}$Sr and $^{174}$Yb demonstrate the expected recoil and Doppler temperature limits. The method requires a relatively small number of scattered photons and can potentially lead to phase space densities approaching quantum degeneracy in sub-second timescales.

\end{abstract}

\pacs{}

\keywords{Sisyphus cooling, cold atoms}

\maketitle

\section{Introduction}

Ultracold atomic gases are now routinely used for a broad range of science including precision measurements
\cite{Clade2009,Peters1999,Martin2007}, quantum degeneracy studies \cite{BEC95,BEC295,DFG99}, and quantum information applications \cite{EntanglRydberg2010,CNOT2010}. The standard procedure to produce such gases relies critically on laser cooling where the fundamental energy loss unit corresponds to the momentum recoil from photon absorption. The magneto-optical trap (MOT) arrangement for laser cooling can produce temperatures of hundreds of microKelvins or even lower. Here, the combination of weak confinement and density dependent losses from photon-assisted collisions and radiation trapping restricts atomic densities to less than about $10^{11}$ cm$^{-3}$ \cite{Ketterle1993}. For applications requiring higher phase-space densities, evaporative cooling in magnetic or optical dipole traps (ODT) is used. While successful, evaporative cooling suffers from inherent atom loss and demands long trap lifetimes usually on the order of a minute. Most measurements with ultracold atoms can benefit from larger atom numbers and phase-space densities, as well as reduced experimental cycling times.

\begin{figure}[ht]
\includegraphics[width=86mm]{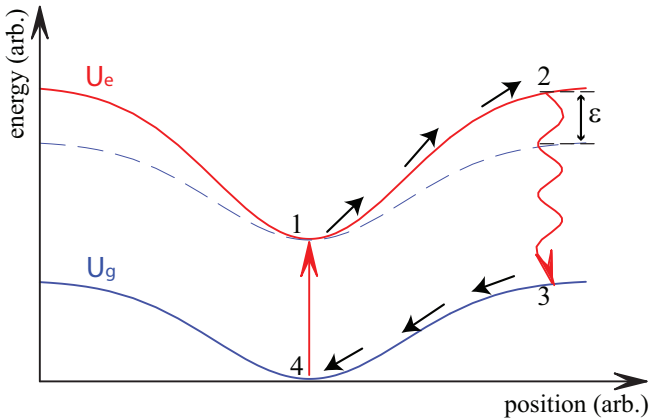}\rule{5mm}{0pt}
\caption{\label{fig:Scheme} (color online) Cooling scheme for a two-state atom in an ODT. An additional cooling beam, near resonant with the atomic transition, is also directed at the atoms. The solid upper red (lower blue) curve shows the spatially-varying ac Stark shift potential $U_e(U_g)$ of the excited (ground) atomic state induced by the trapping beam. The excited state polarizability is greater than that of the ground state. The dashed blue curve is the ground state potential shifted up by the cooling photon energy. The frequency of the cooling beam is set to be resonant with the atom at ODT bottom. The cooling proceeds via the following cycle: 1. a moving atom absorbs a photon at the bottom of the ODT; 2. it climbs a steeper potential (solid upper red curve); 3. it spontaneously decays into the ground state with the shallower potential (solid lower blue curve). The atom loses energy $ \varepsilon$ due to this process; 4. after a number of oscillations, the atom absorbs another photon near the bottom of the ODT and the cycle repeats.}
\end{figure}

In this paper we propose a method of Sisyphus cooling of atoms confined in an ODT that exploits differential ac
Stark shifts of two electronic states. Our method promises phase-space densities which are much higher than those typically achievable with conventional laser cooling and in timescales which are much shorter than those typical of evaporative cooling. In particular, our scheme is well suited for rapid cooling of optically trapped spin-zero atoms which possess narrow electronic intercombination transitions.

In Sisyphus cooling the fundamental energy loss unit is the energy difference between two coupled atomic states at the positions of excitation and de-excitation, and can in principle be very large. Under suitable conditions, the atom is
forced to repeatedly ``climb uphill" and thus lose kinetic energy. The first such proposal by D. Pritchard
\cite{Pritchard1983} involved two rf-coupled hyperfine states in a magnetic trap. Since then this idea has been investigated for a number of atom trap configurations.

The original Pritchard scheme was theoretically studied for the case of Ioffe-Pritchard traps \cite{Bigelow2005} and
modified to incorporate different magnetic energies of two electronic levels \cite{Hoeling1993}. Gravitational Sisyphus cooling in a magnetic trap \cite{Newbury1995} and Sisyphus cooling in a blue detuned evanescent wave trap \cite{Ovchinnikov1997} have been demonstrated. RF-induced Sisyphus cooling in an ODT was demonstrated \cite{Miller2002} and a variant which exploits the second order Zeeman effect and spin-exchange collisions has been proposed \cite{Ferrari2001}. Recently a three-level cooling scheme applicable to magnetically trapped Hydrogen has also been proposed \cite{Porto2011}. All of these schemes rely on transitions between internal magnetic sub-levels of the atom. None of these schemes is applicable to atoms with spin-zero ground state such as strontium or ytterbium.

The rest of this paper is organized as follows. In section II we discuss the general requirements for our cooling scheme and determine suitable ranges of ODT wavelengths for the application of our scheme to Sr and Yb atoms. In section III we develop a simple theoretical model to determine the efficiency of our cooling method and discuss its limitations. In section IV we develop a numerical model that properly addresses the stochastic nature of the process. We use this model to compute the expected value of final temperatures for Sr and Yb. In section V we discuss possible applications of our cooling scheme and draw our concluding remarks in section VI.

\section{Cooling Scheme}

Our basic cooling scheme is shown in Fig.~\ref{fig:Scheme}. A two-state atom confined in the optical dipole potential of the trapping laser interacts with cooling laser light that resonantly couples the internal states near the trap bottom. A moving atom initially in the ground-state absorbs a cooling photon at the bottom of the ODT. The atom in the excited state climbs a steeper potential, thus spending additional energy. Then it spontaneously decays into the ground state far from the minimum of the potential. After a number of oscillations the atom absorbs another cooling photon and the cycle repeats.

Far off resonant optical dipole traps are essentially conservative and atoms will scatter photons only from the cooling laser beam. Elastic collisions between atoms in the ground state are not required for our cooling method but are also not harmful unless the atomic sample is in the hydrodynamic regime. Even though the dynamics of our proposed cooling process are very different from that of a conventional MOT, the two share the same density limiting mechanisms - inelastic light-assisted collisions \cite{JulienneCollisions99} and reabsorption of scattered photons. Crucially, we expect larger equilibrium densities than that achievable in MOTs due to the small excited state fraction as well as the fast cooling timescale. Together with the stronger spatial confinement, the scheme potentially allows far greater phase space densities than those achievable by conventional laser cooling.

\subsection{General requirements}

In order to implement the proposed scheme, one needs to satisfy the following conditions: (1) the ODT wavelength has to provide a higher polarizability for the atom in the excited state than in the ground state, (2) absorption of the cooling light should occur only near the trap bottom, and (3) atoms in the excited state should have enough time to move substantially far from the minimum of the ODT potential (ideally until turning point) to make the cooling process efficient, i.e. the natural decay time should be comparable to the trap oscillation period for the excited state.

We consider the common case of loading of an optical trap directly from a MOT. Condition (1) can be satisfied by proper choice of the ODT wavelength. Condition (2) places a stronger restriction on the ODT wavelength, requiring that the relative ac Stark shift of the two states arising from their unequal polarizabilities is much larger than the linewidth of the cooling transition. This allows the cooling laser to be tuned to the atomic resonance only at particular locations in the trap, such as the bottom. Condition (3) implies an ideal natural decay rate comparable to typical ODT trapping frequencies of less than $100\,$kHz. For a particular choice of atom, conditions (1) and (2) may be satisfied with an appropriate choice of ODT wavelength. Condition (3) however mandates the availability of a narrow cooling transition.

Our scheme is therefore well-matched to the case of alkaline-earth-like atoms (such as Ca, Sr, Yb) that possess
closed, spin-forbidden intercombination transitions with narrow linewidths. Ultracold samples of such atoms are of great interest in the context of optical clocks \cite{Sr87clock1,Sr87clock2,Ybclock}, precise tests of fundamental physics \cite{Sr87clock1,Ferrari2006,Alan2011}, quantum computing \cite{YbQuantComp,YbQuantComp2} and quantum simulation \cite{rey2010}. We focus on applying our scheme to the cases of $^{88}$Sr and  $^{174}$Yb and perform relevant calculations for the $^{1}S_0 \rightarrow {^{3}P_1}$ transition (wavelength $\lambda_{\rm Sr(Yb)}=689(556)$nm,  linewidth $\gamma_{\rm Sr(Yb)}=2 \pi \times 7.4 (182)\,$kHz). Laser cooling of these atoms on this transition is well developed. We first identify suitable ODT wavelength ranges that meet the requirements of our proposed scheme.

\subsection{Suitable ODT for Sr and Yb}

\begin{figure}[ht]
\includegraphics[width=70mm]{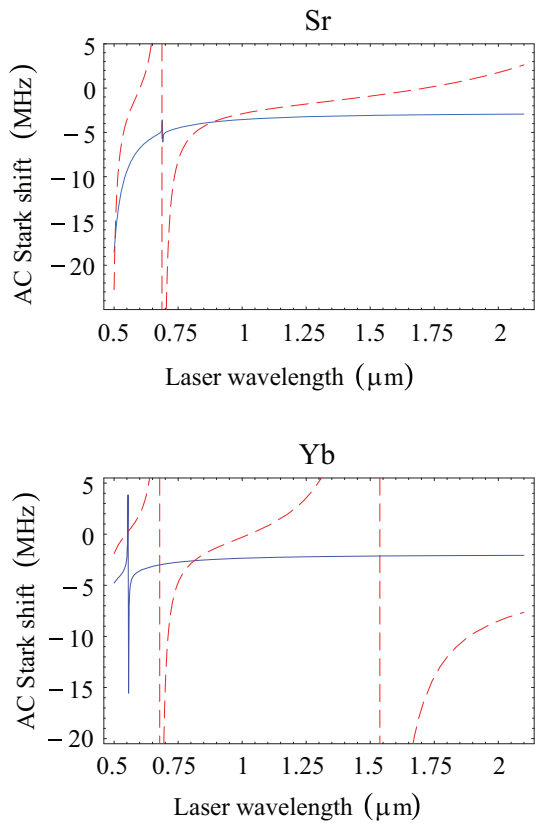}\rule{5mm}{0pt}
\caption{\label{fig:ACStark} (color online) ac Stark shift of the $^{1}S_{0}$ (blue solid line) and $^{3}P_{1}$ (red dashed line) levels versus wavelength for $^{88}$Sr and $^{174}$Yb. For concreteness, the calculation is performed for the maximum intensity in a single 1 W beam focused to a waist of 10 $\mu$m.}
\end{figure}

To determine ODT wavelengths that satisfy the above criteria for Sr and Yb, we calculate the ac Stark shift
(see Fig. \ref{fig:ACStark}) of the $^{1}S_{0}$ and $^{3}P_{1}$ levels for a large range of frequencies detuned to
the red of the strong $^{1}S_0 \rightarrow {^{1}P_1}$ transition.
%We have included all
%contributions of $5s5p ^{1}P_{1},	5s6p ^{1}P_{1}, 	5s7p ^{1}P_{1}$ levels for the shift of $^{1}S_{0}$
% level and $6s5d ^{3}D_{1}, 6s7s ^{3}S_{1},	6s6d ^{3}D_{1}, 6s8s ^{3}S_{1}, 6s7d ^{3}D_{1}, 6p2 ^{3}P_{1}$ levels
 % for the shift of $^{3}P_{1}$ level.
The calculation of the polarizability is performed by summing up the light shift contributions with electronic states up to $n=8$ principal quantum number for both Sr and Yb, using the spectroscopic data presented in \cite{Katori1999}. We have neglected the effect of hyperpolarizability and considered a linearly polarized ODT. For the excited state we assume that atoms are in magnetic sublevels $m=\pm1$ \cite{polfootnote}. The range 700-800 nm appears to be suitable for Sr, since the polarizability of the excited state is substantially higher than that of the ground state. The ranges 690-800 nm and 1700-2500 nm are suitable for Yb. In these regimes, the differential ac Stark shift can be made much larger than the natural linewidth for either atom using modest ODT powers. Hence the cooling beam can be frequency tuned to address atoms only at the minimum of the ODT potential.

\section{Analytical Model}

To gain intuition about the dynamics of the proposed cooling process and understand its limitations, we first develop a 1D analytic model where we neglect the Doppler shift and assume photon absorption strictly at the bottom of the
ODT ($x=0$). These assumptions will be relaxed in subsequent discussions.

In our simple picture, we first estimate the cooling effect from a {\it single} photon scattering event. If an atom
initially in the ground state absorbs a cooling photon at time $t=0$ and goes into the excited state, the probability to still be in the excited state at time $t$ is $P(t)= \exp{(-\gamma t)}$. Here $\gamma$ is the natural decay rate of the excited state or equivalently the linewidth of the cooling transition. The average energy reduction
$\bar{\varepsilon}$ due to a single scattering event within time $t_{d}$ is then expressible as
\begin{equation}
\bar{\varepsilon}=\int_0^{t_{d}} (U_{e}(t)-U_{g}(t)) \gamma \exp(-\gamma t)dt,
\label{eq:coolingrate}
\end{equation}
where $U_{e(g)}(t)$ is the potential energy in the excited (ground) state.

Assuming a harmonic potential $U_{e(g)}=m\omega_{e(g)}^{2}x^{2}/2$, the atom's position before spontaneous emission varies as $x(t)=[(2k_{B}T)/(m\omega_{e}^{2})]^{1/2}\sin(\omega_{e}t)$. Here $\omega_{e(g)}$ is the trap frequency in the excited (ground) state, $m$ is the atomic mass, and $E=k_{B}T$ is the initial energy (kinetic + potential) of the atom. We will be interested in timescales that are much larger than $1/\gamma$. For $t_{d}\rightarrow \infty$ in Eq. \ref{eq:coolingrate}, we obtain
\begin{equation}
\bar{\varepsilon}=k_{B}T \left(1-\frac{\alpha_{g}}{\alpha_{e}}\right)\frac{1}{2}\frac{1}{1+(\gamma/2\omega_{e})^{2}},
\label{eq:coolingrate2}
\end{equation}
where $\alpha_{e(g)}$ is the polarizability of the excited (ground) state. This is the average energy reduction from
scattering a single cooling photon. Here $\alpha_{g}/\alpha_{e}=\omega_{g}^2/\omega_{e}^2<1$.

We can now draw simple conclusions from this intermediate result. The cooling efficiency, i.e. the energy reduction per scattered photon, strongly benefits from high trap frequencies with $\bar{\varepsilon} = 0.5 k_{B}T(1-\alpha_{g}/\alpha_{e})$ in the limit $\omega_{e}\gg\gamma$. However, already at $\omega_{e}=\gamma$, we have an energy reduction of $\bar{\varepsilon} = 0.4 k_{B}T(1-\alpha_{g}/\alpha_{e})$. The energy reduction per transition is proportional to $k_{B}T$, i.e. the full energy of the atom, in contrast to more conventional laser cooling schemes where the energy change per photon absorption is determined by the photon momentum. The proposed cooling scheme demands only a small number of scattered photons, and thus allows the use of transitions with a moderate branching ratio. As will be shown later, substantial cooling can be achieved with few tens of photon scattering events within a fraction of a second.

We now consider two intrinsic heating sources for the atoms. The first is caused by photon recoil which limits the final energy to $k_B T_{\rm rec}=\hbar^{2}{\bf k}^{2}/m$, where ${\bf k}$ is the wavevector of the cooling laser light. The second is caused by the uncertainty in the position at which the atom gets excited. The probability of absorption versus position in the ODT can be written as
\begin{equation}
p(\textbf{r})=A\frac{s}{1+s+\left(\frac{\delta-\delta_{S}
(\textbf{r})-\textbf{k}\cdot \textbf{v}}{\gamma/2}\right)^{2}}.
\label{eq:probabs}
\end{equation}
where $A$ is the normalization constant, $s=I/I_{\rm sat}$ is the saturation parameter (intensity in units of the saturation intensity) of the cooling laser, and $\delta$ is its detuning from the transition frequency of the free atom. $\delta_{S}(\textbf{r})$ is the ODT induced differential ac Stark shift between the excited and ground states and $\textbf{v}$ is the atomic velocity. Even ignoring the Doppler shift, it is impossible to drive atoms into the excited state strictly at the minimum of the ODT potential due to the non-zero linewidth of the transition. This leads to a temperature limitation similar to the Doppler limit for cooling in MOTs, unlike the polarization gradient cooling technique that overcomes the Doppler limit. Neglecting the Doppler shift term and assuming a homogeneous distribution of atoms in the trap we can perform a simple estimate of this heating effect. In 1D, taking $\delta=\delta_{S}(0)$ (i.e. the cooling laser is resonant at the bottom of the trap), the average energy gained during an absorption event is
\begin{equation}
\Delta E=\int_{-\infty}^{\infty} p(x) \frac{m (\omega_{e}^{2}-\omega_{g}^{2})x^{2}}{2} dx
\label{eq:deltaE}
\end{equation}s
which evaluates to $\Delta E = \sqrt{1+s}(\hbar\gamma/2)=k_B T_{\rm Dop}$, exactly corresponding to the Doppler temperature $T_{\rm Dop}$. This can serve as an estimate of achievable temperatures. We can then expect the cooling to cease at either $T_{\rm rec}$ or $T_{\rm Dop}$, whichever is higher.

\begin{figure}[ht]
\includegraphics[width=70mm]{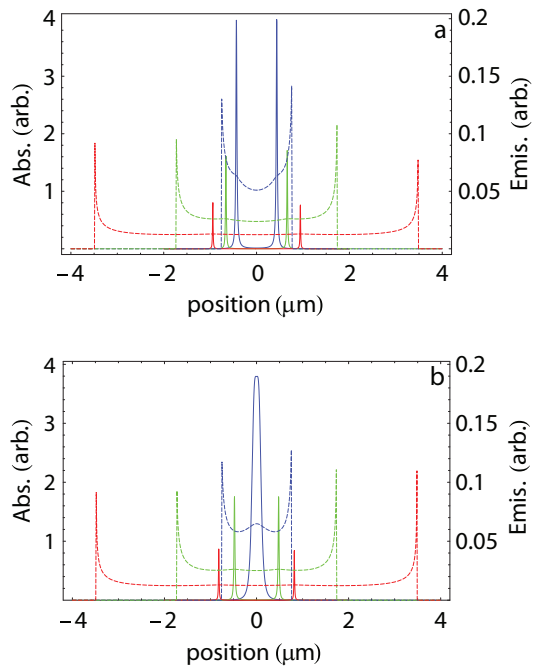}\rule{2mm}{0pt}
\caption{(color online) Absorption (solid lines) and subsequent emission (dashed lines) probabilities of cooling photons per unit length versus position in ODT (centered at position zero). The calculations are performed for Sr atoms moving along the radial dimension of an ODT formed at the 10 $\mu$m focus of a 10 W beam at 750 nm. Curve normalizations take into account that the time spent by an atom at a certain position is inversely proportional to its velocity. The integral under each absorption curve directly corresponds to the probability to absorb a photon during one oscillation. Emission curves are proportional to the rate of emission. The curves correspond to atomic energies of 400 (red (outer) curves), 50 (green (middle) curves), and 5 (blue (inner) curves) $\mu$K and cooling laser detunings of (a) $\delta = \delta_S(0)$  and (b) $\delta = \delta_S(0)-12\gamma$.
\label{fig:Detuning}}
\end{figure}

We now include the Doppler shift due to atomic motion to evaluate the spatial dependence of transition probabilities due to the cooling laser. Fig.\ref{fig:Detuning} shows representative examples of photon absorption and subsequent emission probabilities as functions of position for various energies for Sr. The main consequence of the Doppler effect is to shift the position of peak absorption away from the trap center (see Fig. \ref{fig:Detuning}(a)). This is because while being substantially smaller than the ac Stark shift of the atomic levels, the Doppler shift can easily be larger than the linewidth of the cooling transition ($^{1}S_{0} \rightarrow {^{3}P_{1}}$). Even though the position shift is small compared to the trap size, the cooling efficiency will be reduced. We also note that the effect is far less pronounced in Yb because of its wider cooling transition.

A possible strategy to overcome the effect of the Doppler shift is to introduce an additional detuning of the cooling
light (see Fig.$\,$\ref{fig:Detuning}(b)). In this way the cooling light can be resonant for the atoms near the bottom of the trap. Further refinements include implementing a combination of two or more cooling beams at different frequencies or a frequency chirp of the cooling beam to follow the changing Doppler shift as the atom is being cooled.

In the limit of trap frequencies $\omega_{e}$ much higher than spontaneous decay rates $\gamma$ ($\omega_{e} \gg \gamma$) the emission probabilities will be peaked near the turning points of motion, thus much further from the trap bottom than the absorption probabilities and cooling can be very efficient. When $\omega_{e} \ll \gamma$ the profile of the emission probability is essentially the same as that of the absorption probability and efficient cooling is impossible. When $\omega_{e}$ is comparable to $\gamma$ (as in Fig.$\,$\ref{fig:Detuning}), the emission probability is still greatest near the motional turning points and efficient cooling is possible. This is the regime of focus in this paper.

While the above discussion has been restricted to a single trapped atom, we now consider inelastic collisions in a many-atom system. We can make a simple estimate of the density limitations of this effect by balancing the growth in density from the cooling and the loss in density from two-body inelastic collisions. Atom density in a harmonic trap scales as $T^{-3/2}$. Thus the rate at which density grows due to the Sisyphus cooling is $dn/dt=(dn/dT)(dT/dt)=(-3n/2T)(1/k_B)(dE/dt)$. For an energy loss rate proportional to the energy, we can take $dE/dt=-E/\tau$ and $\tau=(k_BT)/(f\gamma\bar{\varepsilon})$ where $f$ is the average fraction of time spent in the excited state and $\bar{\varepsilon}(\propto k_B T)$ is the average energy loss per photon scattering event. Hyperfine-changing collisions are non-existent for spin-zero atoms making light-assisted collisions the dominant inelastic process. This process leads to a density decay rate $-2\beta f n^2$ where $\beta$ corresponds to the two-body loss rate when $f=1/2$. Our $\beta$ is analogous to that used to characterize light-assisted collisions in bright MOTs where $f \simeq 1/2$ \cite{RaabMOT1987,JulienneCollisions99}. The density reaches equilibrium when the two rates are equal, giving $n_{\rm eq}=3/(4\tau f\beta)$. In the simple situation described by Eq.\ref{eq:coolingrate2}, $n_{\rm eq} = (3\gamma/8\beta)(1-\alpha_g/\alpha_e)/(1+(\gamma/2\omega_e)^2)$. We will make quantitative estimates of $\tau$ for our scheme and the consequent $n_{\rm eq}$ in Section V.

\section{Numerical Model}

In order to address the stochastic nature of the process and to include the effect of the Doppler shift we develop a semiclassical Monte-Carlo simulation. Our numerical model is applied to the dynamics of a single trapped atom and is based on the fact that for a small enough time increment, changes in the relevant parameters such as atomic position and velocity are negligible. We break atomic motion in the ODT into a set of discrete time-steps $\Delta t$, and at each step calculate the probability of the atom to make a transition (either absorption or spontaneous emission). We calculate the new position and velocity assuming that the acceleration is constant during $\Delta t$. If a transition occurs during a particular step, the atom starts the next step in the other state and experiences the different trapping potential. $\Delta t$ is chosen to be much smaller than all other timescales of the problem (trapping frequencies and spontaneous decay rate) and is $1 \mu$s or less for all the calculations in this paper. We include the photon recoil but neglect atom-atom interactions in the simulation.

For an atom in the ground state, we model the absorption of a cooling photon as a random process that happens with
probability given by the right-hand-side of Eq. \ref{eq:probabs} with $A=(\gamma/2)\Delta t$. We set $\delta=\delta_{S}({\bf 0})+\delta_{D}$, where $\delta_{D}=0$ brings the laser into resonance with a stationary atom at the trap bottom and a finite $\delta_{D}$ can be used to mitigate the effect of the Doppler shift. We model the spontaneous decay of an atom in the excited state as a random process with probability defined by the spontaneous decay rate.

\subsection{1D case}

\begin{figure}[ht]
\includegraphics[width=70mm]{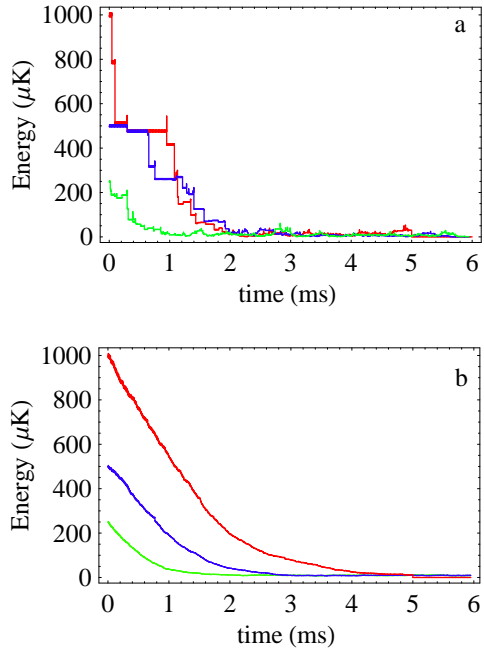}\rule{5mm}{0pt}
\caption{\label{fig:YbSim} (color online) Energy of Yb atoms versus cooling time. Here $\Delta t = 0.1 \mu$s. Harmonic potentials are assumed with trap frequencies $\omega_{g(e)}= 2\pi \times 55(74)$ kHz (see text). The cooling beam parameters are $s=0.5$ and $\delta_{D}=-\gamma$. (a) shows the atomic energy during three individual runs (iterations) of the numerical model. Each curve shows the time evolution for a single initial energy. The sharp steps correspond to transitions from ground to excited state and vice versa. (b) shows the results for the same conditions with each curve averaged over 100 iterations. Average final energy is $\simeq 5.5\,\mu$K in this example.}
\end{figure}

We first employ this numerical model for Yb in a 1D harmonic potential (see Fig. \ref{fig:YbSim}). We use trap frequencies corresponding to the center of the transverse profile at the focus of a 750 nm ODT with 2 W power and $3$ $\mu$m waist. In Fig. \ref{fig:YbSim}(a) we present the evolution of the energy of the trapped atom. The discrete steps up or down in energy correspond to events of absorption or emission of a photon. The exact shape of the curve and the final energy varies between different numerical iterations due to the stochastic nature of the process. In Fig. \ref{fig:YbSim}(b), we show the average of 100 iterations with the same initial conditions. The energy drops to 5.5 $\mu$K within the first 3 ms, comparable to $T_{\rm Dop}=4.3$ $\mu$K, and then stays essentially the same. Here and in the rest of the paper we will express the energy of the atom in $\mu$K.

\begin{figure}[ht]
\includegraphics[width=70mm]{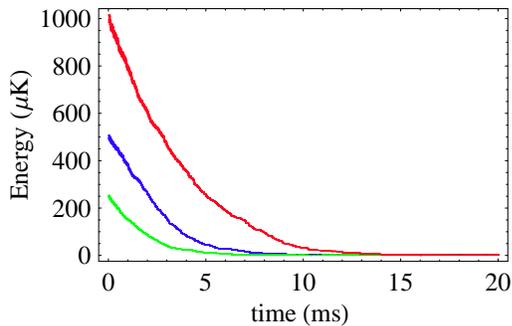}\rule{5mm}{0pt}
\caption{\label{fig:SrSim} (color online) Energy of Sr atoms versus cooling time. Each curve shows the evolution for a single initial energy and is averaged over 50 iterations. Here $\Delta t = 1 \mu$s. Harmonic potentials are assumed with trap frequencies $\omega_{g(e)}= 2\pi \times 10(14)$ kHz (see text). The intensity of the cooling beam is $s=10$ equally split between frequency components corresponding to $\delta_{D}=-\{5,15,45\}\gamma$. Average final energy is $\simeq 0.86\,\mu$K in this example.}
\end{figure}

Similar simulations can be performed for Sr (see Fig. \ref{fig:SrSim}). The smaller linewidth of the cooling transition leads to lower Doppler temperature but also a longer cooling timescale. The harmonic potentials correspond to the transverse profile at the focus of a 750 nm ODT with 5 W power and $10$ $\mu$m waist. To mitigate the more severe Doppler effect due to the narrower linewidth of the cooling transition (see Fig.\ref{fig:Detuning}), we introduce 3 frequency components in the cooling beam. Energies of about 1 $\mu$K are reached within 15 ms.

\begin{figure}[ht]
\includegraphics[width=70mm]{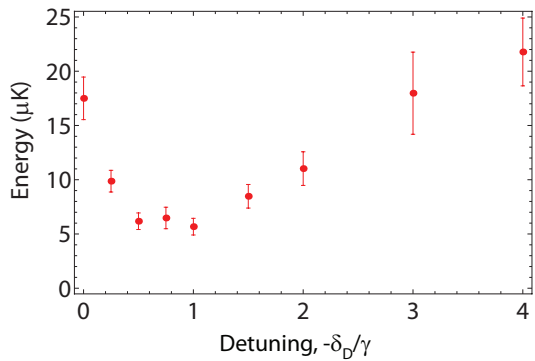}\rule{5mm}{0pt}
\caption{\label{fig:Tfvsdet} (color online) Final energy of a Yb atom versus cooling beam detuning expressed as $-\delta_D/\gamma$. Here $\Delta t = 0.1\,\mu$s, $s=0.5$, and initial energy is $500$ $\mu$K. The final energy is evaluated as the energy after $20\,$ms, averaged over 100 iterations. Each point is the mean and uncertainty of four such values.}
\end{figure}

We studied the dependence of final energy on detuning (Fig. \ref{fig:Tfvsdet}) for the Yb case, with an initial energy of 500 $\mu$K. The initial Doppler shift is about 3 $\gamma$. Clearly, very small detunings will not be useful since they cannot mitigate this shift. Detunings that are much larger than the Doppler shift would also clearly not be useful. This would suggest optimum (red) detunings on the scale of $\gamma$, which is verified by our simulations. For the Sr case, a clear study is impeded by the requirement of multiple frequency components for good cooling performance.

\begin{figure}[ht]
\includegraphics[width=70mm]{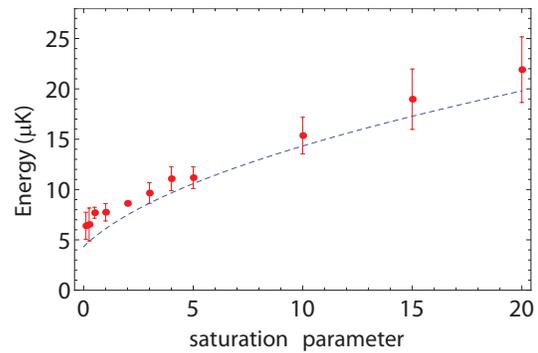}\rule{5mm}{0pt}
\caption{\label{fig:Tfvsint} (color online) Final energy of a Yb atom versus the intensity (expressed as saturation parameter) of the cooling beam. Here $\Delta t = 0.1\,\mu$s, $\delta_D = -\gamma$, and initial energy is $500$ $\mu$K. The final energy is evaluated as the energy after $5\,$ms, averaged over 20 iterations. Each point is the mean and uncertainty of four such values. Also shown is the Doppler limit of conventional laser cooling for a given saturation parameter (blue dashed line).}
\end{figure}

We also investigated the dependence of the final energy on the intensity of the cooling beam. High intensity causes
saturation broadening that makes the position of photon absorption less defined. On the other hand low intensity of
the cooling beam makes events of photon absorption rare, slowing down the cooling process. Therefore, it is meaningless to ask about optimal intensity of the cooling beam without specifying the interaction time. To study this dependence, we set the interaction time to 5 ms. We show in Fig. \ref{fig:Tfvsint} the final energy versus intensity of the cooling beam and compare it to the Doppler limit of $\sqrt{1+s}(\hbar\gamma/2)$.

\subsection{2D and 3D cases}

After gaining some basic understanding we can attempt to apply our numerical method to higher dimensions.
The present cooling method strictly requires that the atom pass through the minimum of the trapping potential. This
will always be the case in 1D. 2 and 3 dimensional cases however allow trajectories that never cross the minimum of the trapping potential. This will make efficient cooling impossible. We verified this in simulations with isotropic 2D or 3D confinement, performed in polar or spherical coordinates, where we observed reduction in the radial velocity component but not in the angular ones. In such a situation, mixing between the different degrees of freedom is desirable. This can be achieved either by means of collisions with other trapped atoms, or by anharmonic mixing. We concentrate our study on the second mechanism, since it does not imply any requirements on collisional properties of atoms or the density of the atomic sample.

Anharmonic mixing is a process that couples atomic motion between different dimensions of a trap. This coupling enables redistribution of energy over the different dimensions. ODT potentials are intrinsically anharmonic. Since atomic energy is not negligible compared to the trap depth, anharmonicity plays an important role.
However mixing between radial and angular components does not occur in spherically symmetrical
potentials. To introduce mixing, an elliptical trap shape is required in the 2D case. In the 3D case, an ODT consisting of two intersecting beams of different transverse sizes can provide the needed mixing. One can expect longer cooling times as well as higher final energies for 3D traps. This is because of the presence of trajectories that do not pass through the center of the ODT, causing the atoms to spend less time at the minimum of the trapping potential. This reduces the excitation rate as well as the average energy lost per scattering event, thus lengthening the cooling process. Additionally, the probability for atoms to be driven into the excited state away from the potential minimum is higher than in the case of 1D traps. This increases the average energy gained per excitation and limits the final energies to values higher than in 1D (see Eq. \ref{eq:deltaE}).

\begin{figure}[ht]
\includegraphics[width=70mm]{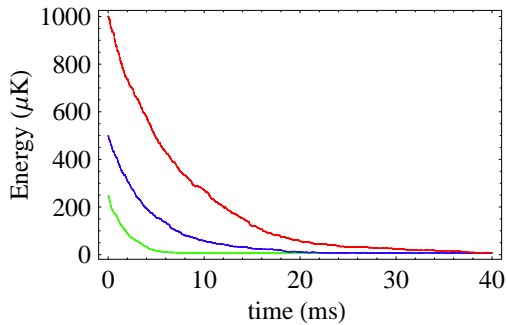}\rule{5mm}{0pt}
\caption{\label{fig:finalplot} (color online) $1/3$ of total energy of Yb atoms versus cooling time, evaluated in 3D simulations. Each curve shows the evolution for a single initial energy and is averaged over 50 iterations. Here $\Delta t=0.1\,\mu$s. The atoms move in an anharmonic crossed-optical dipole trap potential, as described in the text. The cooling beams have detuning $\delta_D=-\gamma$ and total intensity $s=0.5$, split into three equal parts aligned orthogonal to each other with two along the ODT beam axes. The average final energy after 50 ms of cooling is $\simeq 3\times8.6\,\mu$K.}
\end{figure}

We consider two $2\,$W ODT beams with wavelength $750\,$nm, focused to waists of 3 and 3.5 $\mu$m intersecting perpendicularly at their foci. Fig. \ref{fig:finalplot} shows an example of cooling of Yb in such a trap. While a single cooling beam is sufficient to lower atomic energy substantially, our simulations indicate that 2 or 3 beams \cite{polfootnote} will provide faster cooling and about 30$\%$ lower final energies. Employing 3 orthogonal cooling beams at total intensity $s=0.5$ allows energy decrease from $3\times500\,\mu$K to $\simeq3\times10\,\mu$K during 40 ms of cooling from scattering as few as 20-40 photons. As expected, 3D cooling takes a considerably longer time than in 1D. We write the energy as 1$k_{B}T$ per dimension. This definition of $T$ then directly corresponds to the temperature for a sample of many atoms. While thermal equilibrium is not guaranteed by our scheme, equilibrium will be established within a few collisional timescales after the cooling process.

We summarize our findings in Table I where we report the final energies for Sr and Yb for 1D and 3D simulations.
The Sr energies (10x12$\mu$m crossed ODT with 5W in each beam) were obtained for the saturation parameter $s=10$ per cooling beam. Lower final energies could be obtained at the cost of longer cooling time at lower intensities. Our results compare well with the limits set by the Doppler and recoil temperatures.

\begin{table}[ht]
\centerline{
\begin{tabular}{|c|c|c|c|c|c|c|c|c|c|c|c|c|}
\hline
& $T_{\rm rec}$ &  $T_{\rm Dop}$ &  $E_{1D}$&  $E_{3D}$\\  \hline
Sr & 0.46 & 0.18 & 0.8 & $3\times1.4$ \\  \hline
Yb & 0.35 &  4.3 & 6.0 & $3\times 8.5$ \\  \hline
\end{tabular}}
\caption{ \label{fig:Table 1} The final energies (in $\mu$K) for Sr and Yb obtained in 1 and 3 dimensional simulations. Each value of final energy is the result of 80 time averaged numerical simulations.}
\end{table}

\section{Possible Applications}

Our proposed method can potentially enhance many experiments based on optically trapped atoms, including studies
with quantum gases, precision metrology, and quantum information science. The method can act as an additional cooling step that improves initial, pre-evaporation conditions for these experiments. For properly chosen parameters, it may be possible to increase the phase space density to approach quantum degeneracy. The achievable phase space densities can be estimated by incorporating the limitations from density-dependent loss processes. Optimum performance will require a careful choice of ODT and cooling beam parameters. It is experimentally easier to implement the proposed cooling method in ODTs with a larger volume. However a sufficiently high optical power at given wavelength is not always available. We perform calculations for the cases of $^{174}$Yb atoms in 2000 nm wavelength ODT and $^{88}$Sr in 800 nm wavelength ODT. High power at these wavelengths can be obtained from commercially available lasers.

We first examine the case of $^{174}$Yb atoms in an ODT formed by the perpendicular crossing of beams with waists 10 and 12 $\mu$m, each with 20 W power. For an initial energy of $3 \times 500\,\mu$K, our one-atom numerical simulations indicate an average final energy of $\simeq 3 \times 9\,\mu$K using a single cooling beam with $s=3$ and $\delta_D=-0.5\gamma$. The trap depth (5mK) and volume seem adequate to efficiently capture atoms from a MOT operating on the ${^{1}S_{0}}\rightarrow{^{3}P_{1}}$ transition.

We can estimate the density limitations from light-assisted collisions using the expression $n_{\rm eq}=3/(4\tau f\beta)$ derived in Section III. Here we will assume that this is the dominant limiting effect. For simulations with initial energy $0.5\,$ mK, we obtain average values of $\tau = 65\,$ms and $f=0.0025$. Using $\beta = 10^{-11}$cm$^{-3}$/s \cite{betafootnote}, we then get $n_{\rm eq} \simeq 10^{15}$ cm$^{-3}$ as the limiting density. The resultant phase space density at a temperature of $9$ $\mu$K is 0.05, which greatly exceeds what is typically achievable in optically trapped Yb, prior to evaporative cooling \cite{Takasu2003,Ivanov2011}. While the estimated final optical density near the center of the trap is about 6, radiation trapping of the emitted photons should not be an issue, because emissions take place preferentially away from the center, at the motional turning points. Furthermore, the short timescale for cooling will alleviate losses from other inelastic processes such as three-body collisions as well. For subsequent studies, the trapped sample lifetime can be extended by adiabatically relaxing the ODT confinement after cooling.

The cooling performance can be improved by using time-dependent cooling beam parameters. Using $^{88}$Sr atoms in an ODT formed by the perpendicular crossing of beams with waists 10 and 12 $\mu$m, each with 10 W power (7$\,$mK depth), our simulations indicate cooling from $3 \times 250\,\mu$K to $\simeq 3 \times 1.15\,\mu$K with three orthogonal cooling beams with parameters $s=0.5+5.0e^{-t/100{\rm ms}}$ and $\delta_D=-\{3,10,20\}\gamma$. These simulations indicate average values of $\tau = 78.1\,$ ms and $f=0.01$, which yield $n_{\rm eq} \simeq 1.9 \times 10^{14}$ cm$^{-3}$ using $\beta = 10^{-11}$ cm$^{-3}$/s \cite{betafootnote} and a consequent final phase space density of 0.5.
This is of particular significance as direct evaporative cooling of $^{88}$Sr is infeasible due to the near-zero atomic collision cross-section. The density limiting mechanisms may also be further mitigated by using time varying cooling beam detunings and relaxing the ODT confinement near the end of the cooling process. We also note that the spontaneous scattering rate of ODT photons for the above cases are $0.3\,$s$^{-1}$ (Yb) and $18\,$s$^{-1}$ (Sr). The large value for the Sr case may start to affect achievable temperature and phase-space density.

We now turn to applications of this scheme beyond the wavelength ranges and atomic species discussed thus far. We first note that the use of time-dependent cooling beam parameters could allow extension of the proposed method to optical trap wavelengths satisfying $\alpha_{g}/\alpha_{e}>1$. In this case, $\delta$ has to be increased over time, starting from $\delta \simeq 0$.

In addition to alkaline-earth-like atoms, this method can be employed for other atomic species as well, provided sufficiently narrow cycling transitions are available. In alkali atoms, while the usual D2 line $nS_{1/2}\longrightarrow nP_{3/2}$ will be much too broad, the narrower $nS_{1/2}\longrightarrow (n+1)P_{3/2}$ line may be utilized. This transition has recently been used for conventional laser cooling of $^{6}$Li\cite{Hulet2011} and $^{40}$K\cite{Thywissen2011}.

Application of this method to spin-polarized Fermi gases can prove a useful alternative to sympathetic cooling strategies. In this case an additional advantage is derived from the Pauli suppression of inelastic collisions between ground state atoms. A particularly appealing application may be in the context of an optical lattice trap where only a few atoms are located in each lattice site. Atoms in different sites will then enjoy high trap frequencies and thus efficient cooling, while at the same time the total number of traps and therefore atoms can be substantial.

\section{Conclusions}

We have described a Sisyphus cooling method for atoms confined in an optical dipole trap that exploits the trap-induced differential ac Stark shift of two atomic levels coupled through a narrow-linewidth optical transition. With a proper choice of ODT wavelength and energy levels, the atoms are cooled by a process of preferential absorption of photons from a resonant laser beam near the trap center followed by spontaneous emission away from the center. The presented cooling scheme resembles other Sisyphus cooling  methods but doesn't rely on the presence of ground state magnetic sub-structure and is well suited for cooling alkaline-earth-like atoms which have narrow intercombination transitions.

Numerical simulations were presented for $^{88}$Sr and $^{174}$Yb which show that temperatures as low as few $\mu$K are reachable in timescales of tens of milliseconds. The temperature is limited to the higher of Doppler and recoil temperatures. We pointed out experimentally accessible parameters for Yb and Sr where our predicted final phase space density from this cooling method is near the quantum degenerate regime. Further improvements include the use of time dependent trap geometries as well as time dependent cooling beam parameters. The scheme is also adaptable to narrow electronic transitions in other atomic species.

The presented cooling method falls under the more general class of one-photon cooling methods where noticeable reduction of the temperature is achieved by scattering of one photon, as demonstrated in \cite{Bannerman2009,Falkenau2011}. These methods are crucial for cooling species that do not possess near-cycling transitions with high branching ratio. An appealing prospect is cooling of molecules that have been pre-cooled using other methods \cite{Doyle1998,Bethlem2000,Narevicius2007}. Recent work in the DeMille group has demonstrated laser cooling of heteronuclear molecules \cite{Shuman2010} on a transition with a moderately high branching ratio, adequate for cooling using the proposed method.

We thank Mark Saffman for helpful discussions and Alan Jamison for a critical reading of the manuscript as well as useful discussions about this subject. This work was supported by the National Science Foundation grant PHY-0906494 and a National Institute of Standards and Technology Precision Measurements Grant. S.G. acknowledges support from the Alfred P. Sloan Foundation.


\begin{thebibliography}{10}
\bibitem{polfootnote}
We note that the requirement of $\sigma$ transitions can be met with arrangments of 1, 2 or 3 orthogonal cooling beams.
\end{thebibliography}
\end{document}